\documentstyle[12pt]{article}
 \title{ $K\bar{K}$ THRESHOLD  PHENOMENA, $f_0-a_0$
 INTERFERENCE AND KAONIUM PRODUCTION}
 \author{B.Kerbikov\\ Institute of Theoretical and
 Experimental Physics\\ Moscow 117259, Russia\\
 S.V.Bashinsky \\
 M.I.T., Cambridge, MA,USA \\}
  \date{}
\newcommand{\be}{\begin{equation}}
\newcommand{\ee}{\end{equation}}
\begin{document}
\maketitle

We develop a new  formalism to  study $K\bar{K}$
threshold phenomena with resonances $f_0$, $a_0$ and $K^+K^-$ atom
(kaonium) included. The approach provides two possible scenarios for
$f_0-a_0$ mixing. Drastic interference patterns can be
observed   in $K^+K^-$electroproduction at CEBAF, $pd\to
~^3HeX$  reaction, and $N\bar N$ annihilation.

\vspace{1cm}

It is well known that there is an interesting sometimes controversial
physics bearing upon the $K\bar{K}$ threshold phenomena
 and the nature of $f_0,$ and $a_0$ resonances. We try to get a
 different  handle of these issues by constructing an effective
 Hamiltonian which describes
  $\bar KK$ threshold phenomena,
  $f_0$ and $ a_0$ resonances, and  kaonium [1,2].

      Consider reaction  of the type $a+b\to X+M\to \pi^+\pi^-+M$
 with  invariant mass of $X$ being in the vicinity of the $
 K\bar{K}$ threshold(s). This might be, e.g., $pd\to
 ~~^3He\pi^+\pi^-$, or $\bar p p \to \pi^+\pi^-\eta$, or
 electroproduction reaction. Since  $f_0$ and $a_0$ mesons play an
 important role in
  $K\bar{K}$ threshold
 region, we write the amplitude  in the following form
 \be
 T=<\hat{V}|\hat{G}|\hat{A}>+T^0,
 \ee
where $T^0$ is the background (not via $f_0$) $\pi\pi$ production
amplitude,
\be
\hat V=(V_f,V_a), \hat A=\left( \begin{array}{l}
A_f\\
A_a\end{array}\right),
\hat G=\left( \begin{array}{ll}
G_{ff}&G_{fa}\\
G_{af}&G_{aa}\end{array}\right)=
(m_X-\hat H-\Delta \hat H)^{-1}.
\ee
Here $A_f$ and $A_a$ are the $f_0$, $a_0$ production amplitudes, the
propagator $G$ includes their mixing (see below), $V_f$ and $V_a$
couple $f_0$ and $a_0$ to other states (note that $a_0\to\pi\pi$ is
forbidden). Finally
\be
\hat H=\left( \begin{array}{ll}
E_{f}^{(0)}&0\\
0&E_{a}^{(0)}\end{array}\right),
\Delta \hat H = \sum_{\alpha}\hat
V^+\frac{|\alpha><\alpha|}{m_X-m_{\alpha}+i0}\hat V,
\ee
where $|\alpha>$ indicates any of the states to which $f_0$ and $a_0$
are coupled.

Now we  are  furnished to provide two possible $f_0-a_0$ mixing
scenarios. One of them has its origin in the $K^0\bar K^0$ and
$K^+K^-$ threshold splitting of 8 MeV. Taking $\Delta \hat H$ as a
sum of two terms corresponding to $K^+K^-$ and $ K^0\bar K^0$
channels we get
$$
\hat H+\Delta \hat H=
\left( \begin{array}{ll}
E_{f}-i\frac{\Gamma_f}{2}&0\\
0&E_{a}-i\frac{\Gamma_a}{2}\end{array}\right)+
$$
$$
+ D
\left( \begin{array}{ll}
1&\zeta\\
\bar {\zeta}&|\zeta|^2\end{array}\right)
(-i) \sqrt{\frac{m_X-2m_{K^+}}{m_K}+i0}~~+
$$
\be
+ D
\left( \begin{array}{ll}
1&-\zeta\\
-\bar {\zeta}&|\zeta|^2\end{array}\right)
(-i) \sqrt{\frac{m_X-2m_{K^0}}{m_K}+i0},
\ee
\be
D=\frac{m^2_K}{4\pi}|<K^+K^-|\hat V|f>|^2, ~~
\zeta=\frac{<K^+K^-|\hat V|a>}
{<K^+K^-|\hat V| f>}.
\ee

Second mixing scenario is due to kaonium production. Kaonium with its
Bohr  radius of 109.6 fm and binding energy of  6.57 keV is formed by
electromagnetic interaction and its wave function has two isospin
components $I=0,1$ thus providing $f_0-a_0$ mixing. As shown in [1,2]
mixing parameter depends on whether  these mesons are $K\bar K$
molecules or genuine quark states.

The two mechanisms of mixing have very different energy scales --  of
few MeV and few keV  correspondingly.  Typical plots are shown in
Figs. 1,2.

\vspace{1cm}

{\Large{\bf Figure  Captions}}\\

Fig. 1.  The typical plots of the matrix element squared for
$f_0-a_0$ mixing due to $\bar K K$ mass splitting.

Fig.2. Same as in Fig. 1. with mixing due to kaonium.


\begin{thebibliography}{99}
\bibitem{1}
B.Kerbikov, Z.Phys. {\bf 353}, 113, 1995.
\bibitem{2}
S.V.Bashinsky and B.Kerbikov,  Preprint  DAPNIA/SPhN 95 35, Saclay
1995, to appear in Yad. Fiz. 1996.
\end{thebibliography}
   \end{document}